# The Behavior of Flexible MIL-53(Al) upon $CH_4$ and $CO_2$ Adsorption

*Anne Boutin*[a]*, François-Xavier Coudert*[b]*, Marie-Anne Springuel-Huet*[c]*, Alexander V. Neimark*[b,d]*,*

*Gérard Férey*[e] *and Alain H. Fuchs*[b]*

a. Chemistry Department, École Normale Supérieure, CNRS and Univ. Pierre et Marie Curie, 24 rue Lhomond, 75231 Paris cedex 05, France.

b. CNRS, Chimie ParisTech and Univ. Pierre et Marie Curie, 11 rue Pierre et Marie Curie, 75005 Paris, France.

c. Laboratoire de Chimie de la Matière Condensée de Paris, Univ. Pierre et Marie Curie, 75252 Paris cedex 05, France

d. Department of Chemical and Biochemical Engineering, Rutgers, The State University of New Jersey, 98 Brett Road, Piscataway, New Jersey 08854-8058, USA.

e. Institut Lavoisier, CNRS and Université de Versailles St-Quentin en Yvelines, 78035 Versailles (France), and Institut Universitaire de France, Paris, France

AUTHOR EMAIL ADDRESS: alain.fuchs@cnrs-dir.fr

ABSTRACT

The use of the osmotic thermodynamic model, combined with a series of methane and carbon dioxide gas adsorption experiments at various temperatures, has allowed to shed some new light on the fascinating phase behavior of flexible MIL-53(Al) metal-organic frameworks. A generic temperature–loading phase diagram has been derived; it is shown that the breathing effect in MIL-53 is a very general phenomenon, which should be observed in a limited temperature range regardless of the guest molecule. In addition, the previously proposed stress model for the structural transitions of MIL-53 is shown to be transferable from xenon to methane adsorption. The stress model also provides a theoretical framework for understanding the existence of **lp/np** phase mixtures at pressures close to the breathing transition pressure, without having to invoke an inhomogeneous distribution of the adsorbate in the porous sample.

## INTRODUCTION

Gas adsorption in porous solids is known to induce elastic deformation, and this is well documented in the literature, dating back to the first experimental evidence of swelling of charcoal by Meehan and Bangham[1,2] in the late 1920's. The induced strain is usually very small, of the order of $10^{-4}$-$10^{-3}$, and this effect has thus often been neglected in the past discussions and modeling studies of adsorption experiments[3].

In the special case in which the adsorbed fluid is confined to spaces of nanoscopic dimensions (the so-called nanoporous solids), experimental data on adsorption deformation of carbons and zeolites were accumulated over the years by the Russian school of Dubinin and his disciples. This was recently collected and summarized by Tvardovskiy[4]. The effect of adsorption deformations in nanoporous solids is not limited to swelling. Adsorption of gases and vapors in zeolites and carbons[4], as well as in porous silicon[5] or low-k films[6] demonstrates a characteristic common trend: at low vapor pressure the system undergoes *contraction*, followed by *swelling* at higher vapor pressure[7].

The MIL-53 (Al or Cr) Metal-Organic-Framework material has recently attracted a lot of attention on account of its enormous flexibility and the occurrence of an oscillation (or "breathing") during adsorption between two distinct conformations called the large-pore phase (**lp**) and the narrow-pore phase[8-11] (**np**) (see Figure 1), which have a remarkable difference in cell volume of up to 40%. At room temperature and in the absence of guest molecules, the **lp** phase is the most stable form. However, in the course of gas adsorption (such as $CO_2$ or $H_2O$), the **lp** phase transforms into the **np** phase at low vapor pressures, and the reverse transformation occurs at higher pressures. The **lp-np** transition can also be induced by the sole effect of temperature in the empty material. A neutron scattering study of bare MIL-53(Al) was performed by Liu *et al*[12] who observed a reversible **lp-np** transition accompanied by a large hysteresis (the np phase being the stable low temperature phase). In a recent work by Denoyel and coworkers[13], the reversible **lp-np** transition was observed in bare MIL-53(Cr) at room temperature by applying an isostatic pressure all around the sample's microcrystals, using a mercury intrusion device. Molecular simulations have also been used in order to investigate the driving force of the breathing phenomenon at a microscopic level[14].

Some of the present authors have recently interpreted the breathing behavior upon gas adsorption in terms of the non-monotonic sorption induced strain described above[15]. At low vapor pressures, the negative stress that causes a cell contraction eventually induces the **lp→np** transition for large enough deformation, while at higher vapor pressure the reverse **np→lp** transition takes place on account of the change in sign of the induced stress. A "stress model" was derived, in which each structural transition was suggested to occur when the adsorption-induced stress reaches a certain critical threshold. This model successfully accounted for the hysteretic behavior of xenon adsorption in MIL-53(Al) at low temperature[15].

This stress model aims at addressing the issue of the mechanism of the structural transitions associated with breathing. It is basically a "thermo-mechanical" model, in which the transitions are supposed to take place near or at the mechanical stability limits of the **lp** and **np** structures respectively.

On the other hand, Coudert *et al*[16-18] have proposed an equilibrium thermodynamic approach (so-called "osmotic thermodynamic model"), that has successfully rationalized the conditions for the occurrence of breathing in flexible MOF's. It was shown that the occurrence of breathing is conditioned by the relative adsorption affinities of the gas for the two host phases, measured by the ratio of the Henry constants $K^{lp}$ / $K^{np}$, and by the intrinsic stability of the two respective framework conformations, characterized by the free energy difference $\Delta F_{host}$ between the **lp** and the **np** phases. The osmotic model is aimed at predicting what would happen at "full thermodynamic equilibrium" (in the so-called thermodynamic limit, *i.e.* at infinite time). This provides a useful guideline for a generic understanding of the phase behavior of flexible MOF's. Obviously, the real life behavior can only be addressed by a combination of the stress and the osmotic models.

In the light of these two models, we report here an investigation of the effect of methane and carbon dioxide adsorption in MIL-53(Al) at various temperatures. Unlike carbon dioxide, methane does not induce breathing transitions at room temperature[11], which is also the case of argon and nitrogen. It was first suggested that apolar species such as methane or noble gases could not induce breathing because of their too low adsorption enthalpies in MIL-53 materials[11]. More recently, a xenon adsorption study in MIL-53(Al) in the temperature range 195–323 K clearly demonstrated the existence of breathing transitions in the measured adsorption isotherms[19]. A temperature–loading phase diagram was derived, and it was predicted that the breathing effect in MIL-53 was a very general phenomenon, which should be observed in a limited temperature range regardless of the guest molecule[19]. The relative enthalpy and entropy of the bare material were also determined, establishing that the stability of the **lp** phase at high temperature is promoted by entropic effects[19]. This was recently confirmed by quantum chemical calculations[20]. The presently reported adsorption study confirms this prediction. The temperature-loading phase diagram is established for methane and carbon dioxide. Finally, the stress model is applied to the $\{CH_4$, MIL-53(Al)$\}$ system at 224 K and is shown to reproduce quite well the observed hysteretic behavior.

## EXPERIMENTAL METHODS

The methane and carbon dioxide (Air Liquide, Alphagaz, N35 and N48) adsorption–desorption isotherms were measured at various temperatures (from 183 to 298 K in the case of $CH_4$ and 200 to 343 K in the case of $CO_2$), using an "Intelligent Gravimetric Analyser" (IGA system) from Hiden Isochema in the pressure range 0–10 bar. The IGA design allows precise computer-control and measurement of mass change, pressure and temperature.

Prior to sorption measurements, the MIL-53(Al) sample (about 45 mg, of the same origin as the one used in ref. 19) was outgassed at 423 K overnight at a pressure of $10^{-6}$ mbar.

Typically, about 15 to 30 data points were measured within 4 to 6 hours for each isotherm. A thermostat with a water/ethyleneglycol bath was used to measure isotherms at temperatures down to 250 K. For lower temperatures, thermo baths were used: liquid nitrogen/acetone (183 K), dry ice/acetone (200 K), dry ice/acetonitrile (220 K).

## THEORETICAL BASIS

The *osmotic thermodynamic model*[16] is based on the so-called osmotic ensemble, which is the appropriate statistical ensemble to describe fluid adsorption in a flexible porous material. In the osmotic thermodynamic ensemble ($N_{host}$, $\mu_{ads}$, $\sigma$, $T$), the control parameters are the number of molecules of the host framework $N_{host}$, the chemical potential of the adsorbed fluid $\mu_{ads}$, the mechanical constraint $\sigma$ exerted on the system (which, in an isotropic system, is simply the external pressure $P$) and the temperature $T$.

For materials exhibiting clear structural transitions between different metastable framework structures (as opposed to the phenomenon of progressive, continuous swelling for instance), we demonstrated in earlier works that the use of an *osmotic sub-ensemble* adequately describes the equilibrium between host structures upon fluid adsorption[16,17]. In this sub-ensemble the system volume V is restricted to a discrete number of values, corresponding to each metastable structure under consideration. For each host structure $i$, the thermodynamic potential $\Omega_{os}^{(i)}$ and configuration integral $Z_{os}^{(i)}$ in the osmotic ensemble are given by the following equations:

$$Z_{os}^{(i)} = \sum_{N}\sum_{q} \exp\left(-\beta U(q) + \beta\mu N - \beta P V_i\right)$$

$$\Omega_{os}^{(i)}(T,P,\mu) = -kT \ln(Z_{os}^{(i)}) = U - TS - \mu N + PV_i$$

This model was successfully applied to understand the presence or absence of breathing effects in MIL-53(Al) upon adsorption of $CO_2$, $CH_4$, linear alkanes, and more recently xenon as well as $CO_2/CH_4$ mixtures at room temperature[16-19]. As in our previous studies, we used Langmuir fits of the experimental isotherms as approximations to the rigid host isotherms in both the **lp** and **np** structures. In our recent study of xenon adsorption on MIL-53(Al)[19], we have used stepwise isotherms at various temperatures to determine the transition enthalpy and entropy of the empty host material, and so the free energy difference between the empty **lp** and the **np** structures. Not unexpectedly, the **lp** form was predicted to be the most stable one at room temperature, while the **np** structure becomes the most stable one below 203 K. One of the advantages of the osmotic thermodynamic model is that it enables to compute equilibrium thermodynamic data for the bare host material using thermodynamic adsorption data only.

The *stress model*[15] relates the stress exerted by the adsorbed molecules on the adsorbent framework with the adsorption isotherm. From the thermodynamic standpoint, the adsorption stress $\sigma_s$ can be quantified by the derivative of the grand thermodynamic potential $\Omega_c$ of the adsorbed phase per unit cell with respect to the unit cell volume $V_c$ at fixed temperature $T$ and adsorbate chemical potential $\mu$[21, 22].

$$\sigma_s(V_c) = -\left(\frac{\partial \Omega_c}{\partial V_c}\right)_{\mu,T} \tag{1}$$

In pores of simple geometry (slit, cylindrical, or spherical shape), the adsorption stress has a simple physical interpretation as the normal to the pore wall component of the stress tensor in the adsorbed phase[23-25]. In anisotropic materials like MOF's, this interpretation is no longer valid, and one needs to introduce tensor quantities. However, the adsorption stress defined by eq. 1 can serve as an overall scalar measure of the magnitude of the adsorption forces acting on the porous framework. The difference between the adsorption stress, $\sigma_s$, and the external pressure represents the so-called solvation or disjoining pressure, $P_s$, which determines the magnitude of framework elastic deformation in terms of the volumetric strain $\varepsilon$ ($\varepsilon = \Delta V_c / V_c$, where $\Delta V_c$ is the variation of the cell volume), assuming the linear Hooke law with an effective framework bulk modulus $\kappa$, $P_s = \sigma_s - p_{ext} = \kappa\varepsilon + \sigma_0$, where $\sigma_0$ is a pre-stress in the reference state, at which the cell volume $V_c$ is defined[21].

The linear elasticity theory describes adsorption-induced deformations of microporous materials like zeolites and activated carbons when the strain is small, typically in fractions of a percent. For breathing MOF's, experiencing structural transitions with volume changes in tens of a percent, the stress-strain linearity should hold only for the stable **lp** and **np** phases. In the vicinity of the transition, the stress-strain relationship becomes necessarily nonlinear, and it may even diverge at the onset of the transition. We hypothesize that the structural transition occurs when the solvation pressure approaches a certain critical stress σ* that the framework cannot resist. The critical stress, $\sigma^*_{lp}$, associated with the **lp→np** transitions should be negative because this transition corresponds to a framework contraction, while the critical stress of the **np →lp** transition $\sigma^*_{np}$ should be positive. As shown earlier[14], this hypothesis explained the hysteretic behavior of structural transitions in breathing MOF's and was consistent with the existence of two breathing transitions in MIL-53 upon xenon adsorption.

## RESULTS AND DISCUSSION

### Methane adsorption–desorption isotherms

Methane adsorption–desorption isotherms on MIL-53 (Al) were measured in the pressure range 0–6 bar, for a variety of temperatures between 183 and 298 K. Figure 2 shows a comparison of the isotherm obtained in this work, at room temperature (298 K), against the isotherm published by Bourrelly *et al*[26] (obtained at 304 K, up to 30 bar). The two sets of data are in very good agreement, with a smooth, type I curve, and no hint of structure breathing. X-ray diffraction measurements reported in ref. 27 confirm that the structure observed for MIL-53 (Al) at 304 K, is that of the large-pore (**lp**) phase, for methane pressure up to 30 bar.

Figure 3 reports six experimental methane adsorption–desorption isotherms at various temperatures (a seventh isotherm was measured at 224 K and is shown elsewhere for the sake of clarity). At 273 K and 250 K, methane adsorption follows reversible type I isotherms similar to room temperature results, indicating a lack of breathing in the pressure range observed. At lower temperatures, however, adsorption and desorption isotherms exhibit steps and hysteresis loops, which can be linked to adsorption-induced structural transitions (breathing). While it was previously demonstrated on the example of xenon that the occurrence of MIL-53(Al) breathing upon gas sorption depends strongly on temperature[19], it is the first time that this breathing is evidenced in the case of methane, a gas which was so far on the short list of gases known not to trigger breathing of MIL-53.

Furthermore, the experimental stepped adsorption and desorption isotherms can be fitted by two partial Langmuir isotherms, shown in Fig. S1 of the Supplementary Information, as solid and dashed lines respectively. For each stepped isotherms, the two Langmuir fits are approximations of the “rigid host” isotherms of the **lp** and **np** phases, *i.e.* the isotherms that would be obtained for the MIL-53 material frozen in the **lp** (respectively **np**) framework structure. These fits are entirely coherent with the expected thermodynamic properties of the two phases: the narrow-pore phase has a lower saturation uptake $N_{max}$ (around 4 molecules per unit cell) than the large pore structure (8 to 11 molec/uc), and a higher affinity for methane, *i.e.* a higher Henry constant $K_{np} > K_{lp}$.

The fitting parameters ($N_{max}$ and $K$) for both phases are reported as functions of temperature in the Supplementary Information section (figs. S2 and S3). Their evolution with temperature is consistent with the analysis previously performed on xenon isotherms[19]. In particular, the adsorption enthalpy of methane in the **lp** phase, calculated from the slope of $\log(K_{lp})$ vs $1/T$, is found to be $\Delta H_{ads,lp} \approx 16$ kJ/mol. This value is in excellent agreement with the experimental results obtained by calorimetry at room temperature ($\Delta H_{ads,lp} \approx 17$ kJ/mol)[26].

Finally, we checked the internal consistency of the fitting procedure by computing the Langmuir equations for the **lp** and **np** phases at 224 K, using the data collected from the six isotherms reported in Figure 3. These fits were compared with the experimental isotherm (shown in Figure S4 of the supplementary information, but not in Figure 3 for sake of presentation clarity). As can be seen, the agreement is excellent. It clearly displays the existence of a wide hysteresis loop in the 1.5–3.5 bar range, which can be ascribed to the higher pressure **np**→**lp** transition upon adsorption (and **lp**→**np** transition upon desorption). In addition, a smaller step can be detected around 0.35 bar, corresponding to the first lower-pressure **lp**→**np** structural transition. This latter phenomenon was also observed in the 213 K isotherm (figure 3), but no such sign of low-pressure transition **lp**→**np** was observed in the 196 and 183 K isotherms, indicating that the initially empty material was in the **np** phase.

## Carbon dioxide adsorption–desorption isotherms

Carbon dioxide adsorption–desorption isotherms on MIL-53 (Al) were measured in the pressure range 0–10 bar, for a set of temperatures between 200 and 343 K.

Figure 4 reports six experimental $CO_2$ adsorption–desorption isotherms at various temperatures. At 343 K, carbon dioxide adsorption follows reversible type I isotherms, indicating a lack of breathing in the experimental pressure range. At lower temperatures, however, adsorption and desorption isotherms exhibit steps and hysteresis loops, which are obviously linked to the breathing phenomenon. As in the case of methane, the $CO_2$ room temperature isotherm is in good agreement with the one previously published by Bourrelly *et al.*[26] (fig. S5 of Supplementary Information).

The experimental stepwise adsorption and desorption isotherms were again fitted by two partial Langmuir isotherms, as described above in the case of methane. The fitting parameters ($N_{max}$ and $K$) for both phases are reported as a function of temperature in the Supplementary Information section (figs. S6 and S7). The adsorption enthalpy of carbon dioxide in the **lp** phase, calculated from the slope of log($K_{lp}$) vs $1/T$, is found to be $\Delta H_{ads,lp} \approx 38$ kJ/mol. This value is in good agreement with the previously published calorimetry and simulation results at room temperature ($\Delta H_{ads,lp} \approx$ 35-37 kJ/mol)[26,28,29].

## Temperature-loading phase diagrams

We used the osmotic thermodynamic model together with the fits performed on the six experimental methane adsorption isotherms, reported in Figure 3, to investigate the full temperature–loading phase diagram of {$CH_4$, MIL-53(Al)}. By solving the osmotic thermodynamic equations numerically, we determined for each temperature, whether breathing occurs and, if so, what the transition pressures are. All the parameters needed to compute this phase diagram are given in Table

1.

The predicted temperature–methane pressure diagram is shown in Figure 5. The **lp** phase was found to be stable at high temperature and again at lower temperature. There is an intermediate **np** phase stability domain for methane pressure lower than a limiting pressure of around 2 bar. This result is reminiscent of the re-entrant behavior observed in some liquid crystals[30]. As noted above, however, the low-temperature stable phase in the absence of methane (zero pressure) is the **np** phase. It is worth noticing that the 224 K experimental data were not included in the calculation of the diagram. The transition pressures (Figure S5 of the Supplementary Information) are nevertheless in very good agreement with the computed phase diagram, and this provides an extra check of the consistency of both the model and the fitting procedure.

This phase diagram is similar to one previously determined in the case of the {Xe, MIL-53(Al)} system[19]. It starts from the equilibrium **np–lp** temperature $T_0$ of 203 K at zero pressure. It is worth noting that this value, obtained applying our model to xenon adsorption[19], is independent of the nature of the adsorbed gas. It falls within the range of the temperature hysteresis observed by neutron scattering on bare MIL-53(Al)[12]. The initial slope of the transition curve is proportional to the logarithm of ($K_{np}/K_{lp}$), the ratio of adsorption affinities in the two structures, and is thus strictly positive since the affinity of the guest adsorbate for the closed form of the framework is higher than for the open form[16,19]. The condition $K_{np}/K_{lp}>1$ thus favors the closed **np** phase, and consequently the phase transition temperature increases with the gas loading (i.e. the stability domain of the **np** phase increases with $P_{CH4}$). At higher temperature, the transition free energy increases, and it becomes more and more difficult to maintain the **np** form as the most stable one. This causes the observed bending of the transition line above ~230 K. For obvious entropy reasons, the **lp** phase eventually becomes more stable at high temperature, regardless of the methane loading. This situation is also true at high pressure. As the adsorbate pressure increases, at any temperature, the **lp** structure eventually becomes more stable than the **np** one because it can accommodate a higher loading of guest molecules. Since the **lp** phase is the most stable one at high enough temperature as well as at high adsorbate pressure, one has to conclude that the stability domain of the **np** phase should be limited in adsorbate pressure ($P_{max}$~ 2 bar in the case of methane and 1.6 bar in the case of xenon, see Figure 7).

The above thermodynamic considerations are very general and obviously not limited to the special case of methane or xenon adsorption. We predicted earlier[19] that the main features of such a phase diagram would hold for any {guest, MIL-53(Al)} system. The condition $K_{np}/K_{lp} > 1$ is expected to hold true for all the simple guest molecules that have been investigated so far. This means that there should be a range of temperatures above the equilibrium **np–lp** transition temperature of the bare MIL-53(Al) material (203 K in our model, subject to an estimated uncertainty of ± 10 K) where the initially empty open structure contracts upon guest molecule adsorption. The fact that this has not been observed in some cases at room temperature simply means that the transition line maximum in this system, $T_{max}$, is below the room temperature. The present findings for methane clearly confirm these predictions.

Using the same procedure, we have computed the {$CO_2$, MIL-53(Al)} diagram, using the experimental adsorption isotherm shown in Figure 4. All the parameters used to compute this phase diagram are given in Table 1. The calculated diagram for $CO_2$ is shown in Figure 6. The overall shape of the diagram confirms the existence of a generic temperature-loading phase diagram, whatever the guest molecule is.

The three phase diagrams obtained in this work are shown on the same graph, for comparison sake, in Figure 7. The difference in the stability domain of the **np** phase in the three different cases can be qualitatively understood as follows. For each system, there is a temperature $T_{max}$ (see Figure 7), above which gas adsorption does not induce the **lp->np** phase transition anymore. Since the driving force for the closure of the **lp** structure is the guest-host interaction, which induces the cell contraction, one may simply write: $kT_{max} \approx \Delta H_{ads,lp}$. The increase in $T_{max}$ in going from $CH_4$ to Xe and $CO_2$ can then simply be explained by the increase in adsorption enthalpy in the **lp** phase (see Table 1). While $P_{max}$ values are close for $CH_4$ and Xe, the value for $CO_2$ is larger by a factor of ~3. This difference can be accounted for by a larger interaction energy between guest molecules in the case of carbon dioxide, due their quadrupole moments. This causes a stronger ordering of the $CO_2$ molecules in the **np** phase and increases the stability domain of this phase.

It must be recalled at this stage that the osmotic model predicts the conditions of thermodynamic stability at full equilibrium and does not take into account hysteresis effects. Hysteresis was systematically encountered in all reported MIL-53 experiments and often leads to some complicated mixtures of phases[31]. In a recent structural study of MIL-53(Fe), Millange *et al*[32] have attributed some of their results to heterogeneous mixtures of crystallites in either open or closed form, depending on their contact with the guest molecules. Such behavior cannot be taken into account in this model, which only describes what would happen in a homogeneous system at full equilibrium. The osmotic model is aimed at describing the *thermodynamics behind the scene*.

## Hysteresis effects and the stress model

We finally examine how our recently published stress model of structural breathing of MOF's can be applied to some of the data at hand. To do so, we consider that the central quantity determining and describing the structural transitions of the material is the adsorption-induced stress, a stimulus which triggers the breathing transitions. The phase transformation of the host structure thus happens at a certain critical stress threshold that the material in a given phase cannot withstand. As a consequence, this model implies that the adsorption–desorption isotherms exhibit hysteresis loops, since the structural transition pressure depends on the stress threshold of the host structure before the transition, rather than on the condition of thermodynamic equilibrium between the phases. A schematic representation of the adsorption stress for both phases and critical stress $\sigma^*_{np}$ and $\sigma^*_{lp}$ determining the structural transitions upon adsorption and desorption is shown in Figure 8.

We applied the stress model to the adsorption of $CH_4$ in MIL-53 (Al) at 224 K. The reasons for choosing this particular temperature are twofold. Firstly, the experimental data at 224 K display both the lower and higher pressure breathing transitions, thus giving us more information to fit (and to check) the parameters of the model. Secondly, this temperature is close to 220 K, for which the stress model was applied to the adsorption of xenon in the same material, enabling a comparison of the behavior of the two adsorbates and giving some insight into the transferability of the model parameters between different fluids.

Table 2 summarizes the parameters of the stress model. The values of the Langmuir equation parameters ($K$ and $N_{max}$) are taken directly from the isotherm fits. The values of the derivatives ($dN_{max}/dV_c$) are approximated by a linear interpolation, i.e. ($dN_{max}/dV_c$) = ($N_{max,lp}$–$N_{max,np}$) /($V_{c,lp}$–$V_{c,np}$) as done for xenon study[15]. Finally the values of the derivatives ($dK/dV_c$), for which no simple approximation is possible, are fitted to reproduce the experimental transition pressure upon adsorption and desorption, along with the values of the critical stress for both phases, $\sigma^*$. Within the imposed constraints, the fitting is successful and the optimal values of ($dK/dV_c$), reported in Table 2, appear to be similar to the values derived for xenon adsorption[15]. In particular, the ratios ($dK/dV_c$)/$K$ in the **lp** phase are remarkably close, which points to a good transferability of the model parameters. Although a more systematic investigation will have to be performed on a large number of adsorbates, the good transferability of parameters confirms the robustness of the model and its physical significance.

The stress model[15] provides a plausible explanation for *both* the hysteresis and the phase mixture effects that have been almost systematically observed experimentally and discussed in the literature[8,11,32-35]. As seen above, the hysteresis effect can be accounted for by the difference in the stress threshold of the host structures at the structural transition, regardless of possible phase mixture effects. On the other hand, the phase coexistence phenomenon deserves a specific comment. From a macroscopic point of view, the Gibbs phase rule predicts a possible phase coexistence at a single pressure only (for a given temperature, at thermodynamic equilibrium). It has been suggested that a heterogeneous mixture of **lp** and **np** could be accounted for by the fact that some crystallites remained out of contact with the external gas. This explanation cannot be ruled out, even though it seems rather unlikely, given the long equilibration times used in the X-ray experiments for instance. It has also been suggested very recently, based on a Xe NMR study, that a single crystallite could accommodate both the **lp** and the shrunken **np** form[36].

The stress model provides a simpler explanation, assuming that the experimental sample of MIL-53 is composed of a large number of crystallites of different sizes. We hypothesize that the size of a given crystallite influences the critical stress that can be withheld in a particular phase. Thus, the thermo-mechanical state of the system depends not only on the macroscopic temperature and pressure, but also on the distribution of stresses $\sigma_s$. Because of this third state parameter, the local expression of the Gibbs phase rule now reads:

$$D = C + 3 - p \qquad (2)$$

where C is the number of components and p is the number of phases. The degree of freedom D is now equal to 2 (instead of 1), for a single component system with two phases in contact. This allows for phase coexistence to take place over a certain range of pressure, at a given temperature.

We interpret the fact that the **lp→np** and **np→lp** structural transitions do not happen abruptly, as steep steps in the isotherms, but are seen as rather smooth transitions (Figure 9), as linked to the presence of phase mixtures in the sample during the breathing transitions. Considering distributions of critical stresses $\sigma_{np}$* and $\sigma_{lp}$*, and choosing a Gaussian curve as the simplest such distribution, $P(\sigma) = \exp(-(\sigma/\delta\sigma)^2)/(2\pi\,\delta\sigma)^{1/2}$, we chose the additional $\delta\sigma_i$* ($\delta\sigma_{lp}$* = 1.2 $10^{-4}$, $\delta\sigma_{np}$* = 2.0 $10^{-3}$) parameters to be compatible with the spread of the experimental adsorption and desorption steps. The resulting model isotherm, presented in Figure 9, shows a remarkable agreement with the experimental data, lending credit to the model's assumptions. The calculated isotherm also helps identify the first, low-pressure, breathing transition around 0.4 bar, indicated by a small hysteresis loop. Furthermore, using only the adsorption data as input, the stress model allows to predict of phase composition curves along both adsorption and desorption branches, which are displayed in Figure 10. These predicted phase-mixture profiles shed light onto the width of the pressure range of breathing transitions, and they could be compared to phase compositions obtained experimentally, e.g. by *in situ* X-ray diffraction techniques, as a further validation of our theoretical model for this phenomenon.

It must be stressed finally that we have addressed here the breathing transitions that take place upon gas adsorption in MIL-53. Whether or not the transition mechanism in absence of adsorbed gases can also be accounted for by the presently developed stress model is still an open question.

## CONCLUSION

The use of the osmotic thermodynamic model[16-19], combined with a series of methane and carbon dioxide gas adsorption experiments at various temperatures, has allowed us to shed some new light on the fascinating phase behavior of the MIL-53(Al) flexible material. We derived a generic temperature–loading phase diagram, and we predict that the breathing effect in MIL-53 is a very general phenomenon, which should be observed in a limited temperature range regardless of the type of guest molecules, since it is expected that the affinity of any adsorbate for the closed **np** form of the framework is always higher than for the open **lp** structure.

The previously proposed stress model for the breathing structural transitions of MIL-53 was shown here to be transferable from xenon to methane adsorption. This is a very encouraging result demonstrating the robustness and physical significance of the model. Work is in progress to derive a more general temperature-dependent stress model. Last but not least, in addition to providing a plausible mechanism for the breathing transitions upon guest adsorption, the stress model also provides a theoretical framework for understanding the existence of **lp/np** phase mixtures at pressures close to the breathing pressure, without assuming an inhomogeneous distribution of the adsorbate in the porous sample. We believe that these very general models should provide useful tools for experimentalists to better understand the soft porous solids behaviors and to choose the most suitable experimental conditions for studies of structural transitions.

**Acknowledgments.** Dr T. Loiseau is gratefully acknowledged for providing us with the MIL-53(Al) sample. A.V.N. acknowledges the ARO grant W911NF-09-1-0242 and the Région Île-de-France for the support via a Blaise Pascal International Research Chair administered by the Fondation de l'École Normale Supérieure.

**Supporting Information Available:** Fits of the experimental adsorption isotherms as well as fitting parameters for both phases as functions of temperature. This material is available free of charge via the internet at https://pubs.acs.org

**References.**

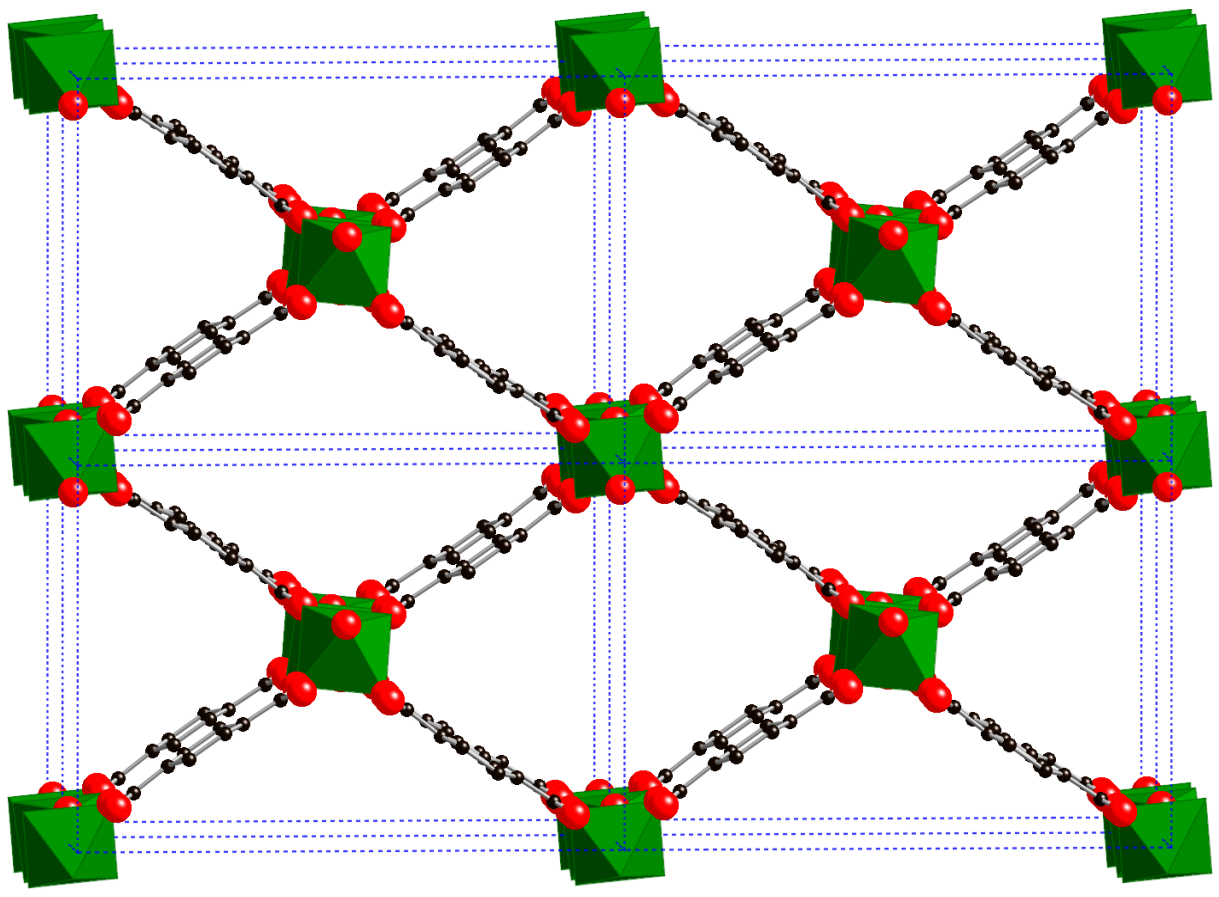

**(lp)**

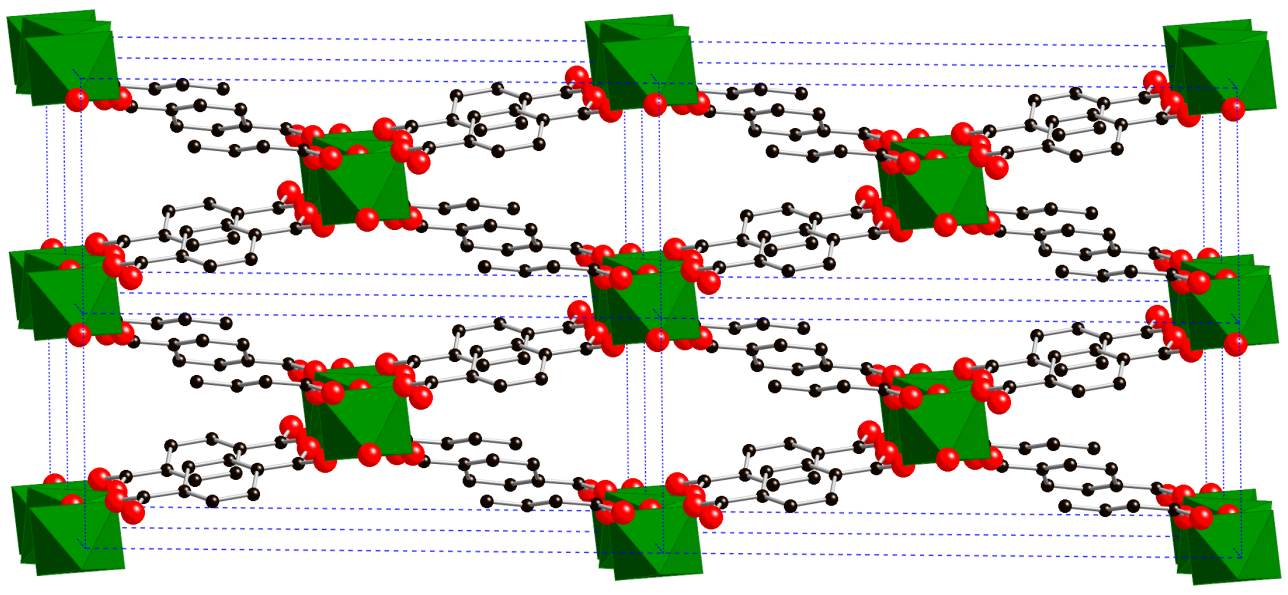

**(np)**

**Figure 1.** Representation of the metastable **lp** and **np** structures of material MIL-53 (Al), as a 2 x 2 x 2 supercell viewed along the axis of the unidimensional channels.

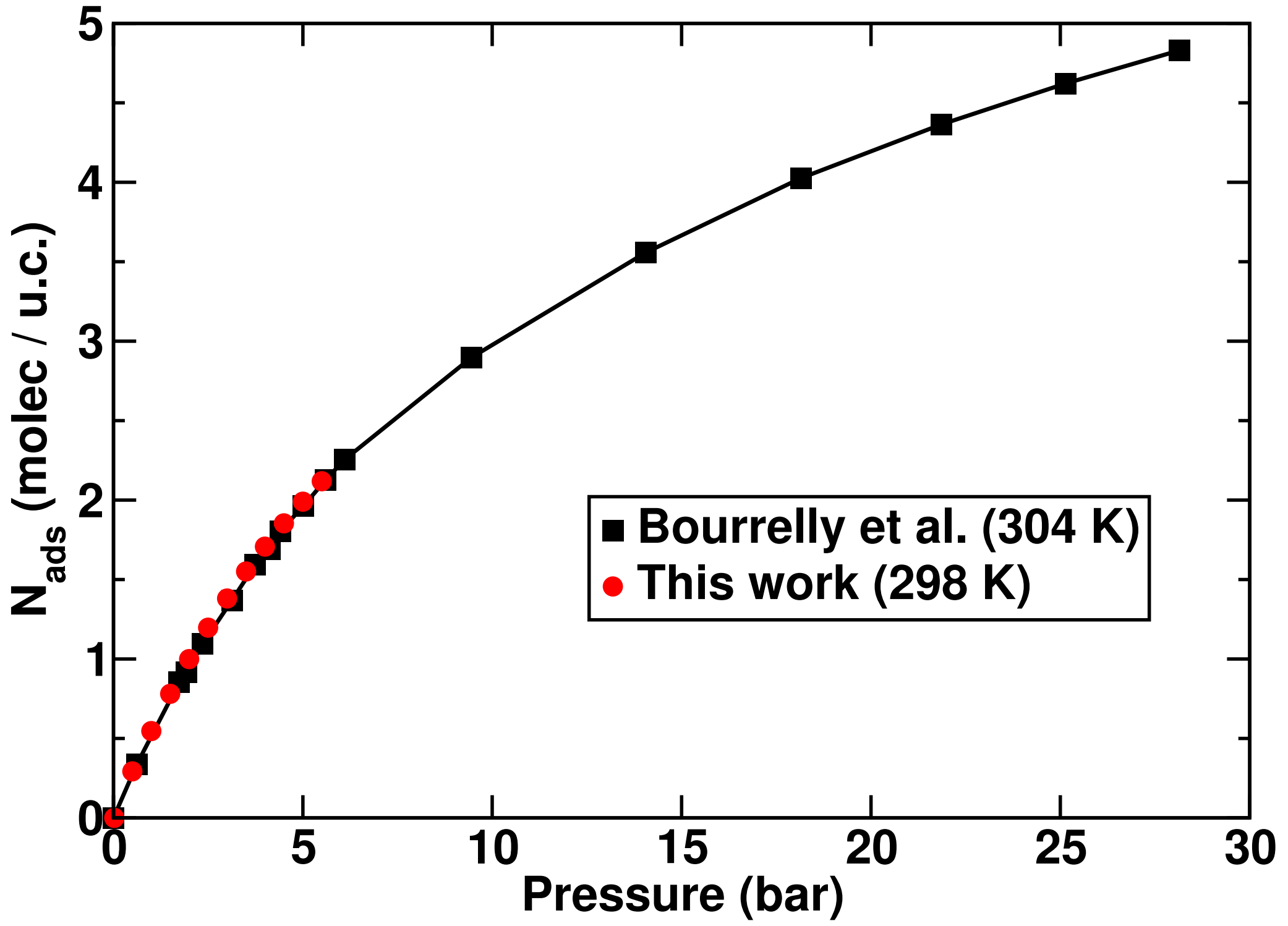

**Figure 2.** Experimental adsorption isotherms of $CH_4$ in MIL-53 (Al) measured at ambient temperature by Bourrelly *et al*[26] and this work.

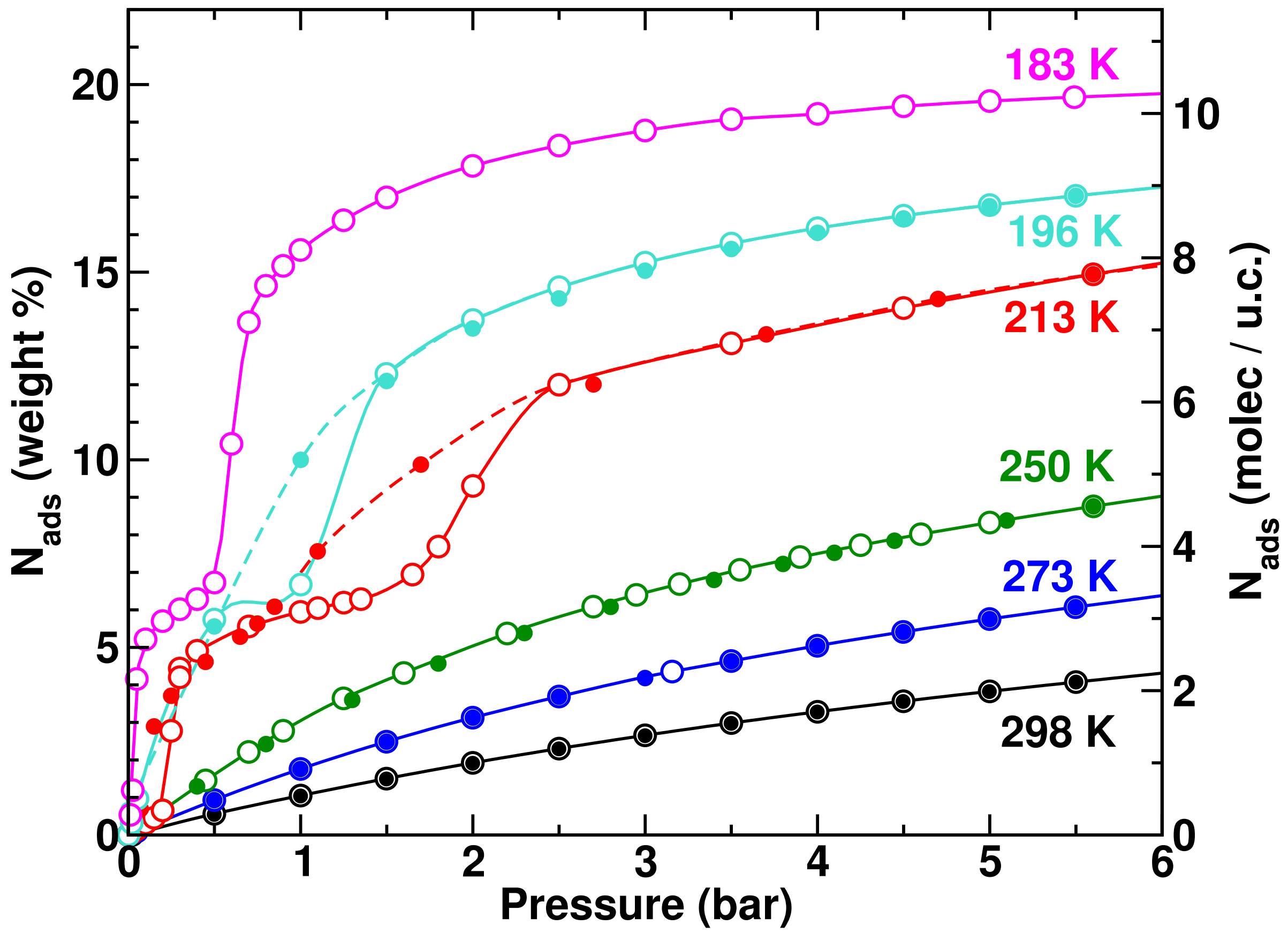

**Figure 3.** Adsorption and desorption isotherms of $CH_4$ in MIL-53 (Al) in the 0–6 bar range for temperatures between 183 K and 298 K. Open symbols: adsorption; full symbols: desorption. The desorption branch at 183 K was not recorded. Lines are guides for the eye.

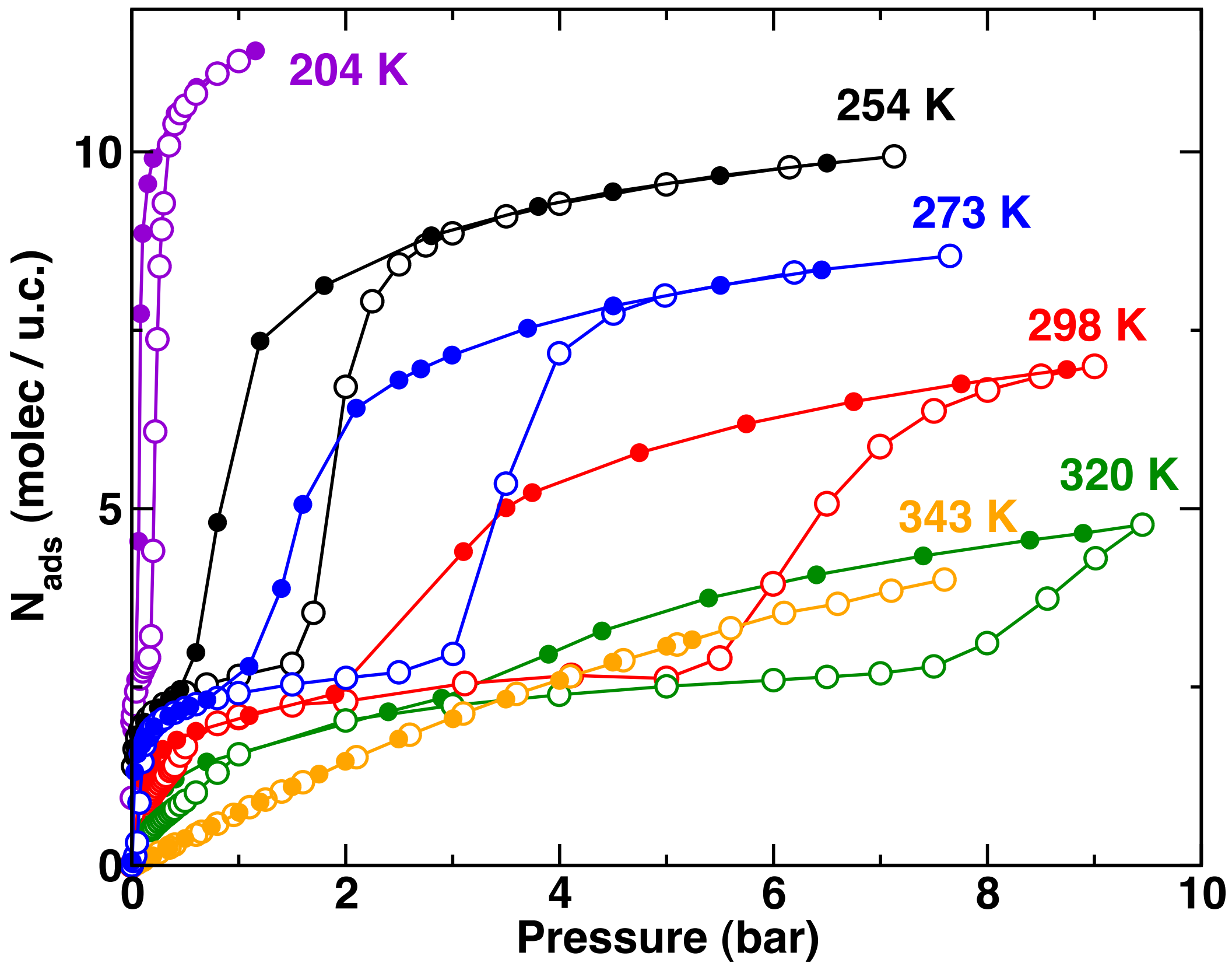

**Figure 4.** Adsorption and desorption isotherms of $CO_2$ in MIL-53 (Al) in the 0–10 bar range for temperatures between 204 and 343 K. Open symbols: adsorption; full symbols: desorption. Lines are guides for the eye.

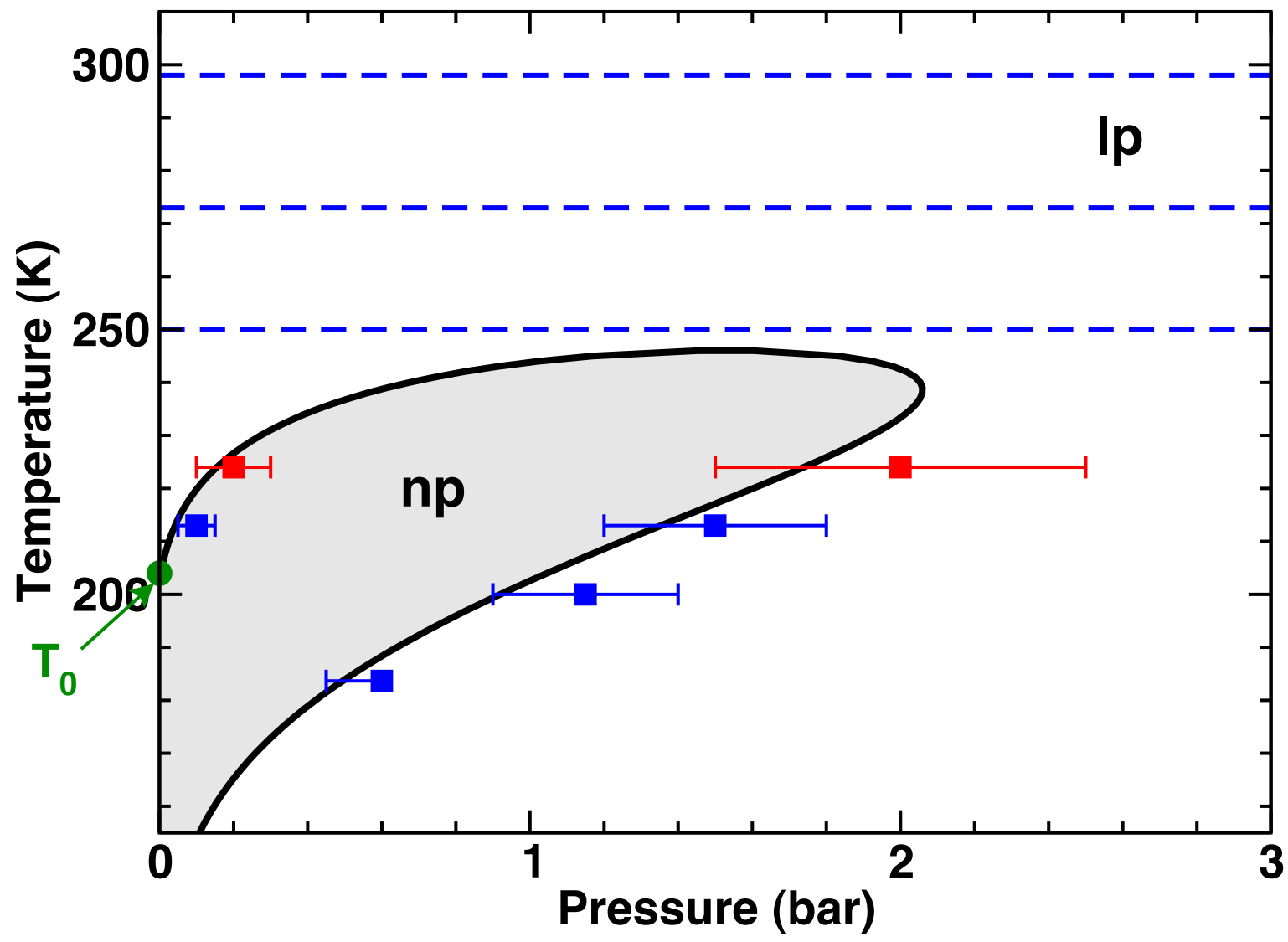

**Figure 5.** Predicted temperature–vapor pressure phase diagram for $CH_4$ adsorption in MIL-53 (Al) material (black line), compared with experimental data points. Blue squares (and error bars) represent the observed structural transitions. The dashed blue lines represent the isotherms (250 K, 273 K and 298 K) for which no transition was experimentally observed in this pressure range. The red symbols (and error bar) correspond to data obtained at 224 K that were not included in the computation of this phase diagram (see text). $T_0$ is the equilibrium **lp-np** phase transition temperature, derived by our model, in the empty material.

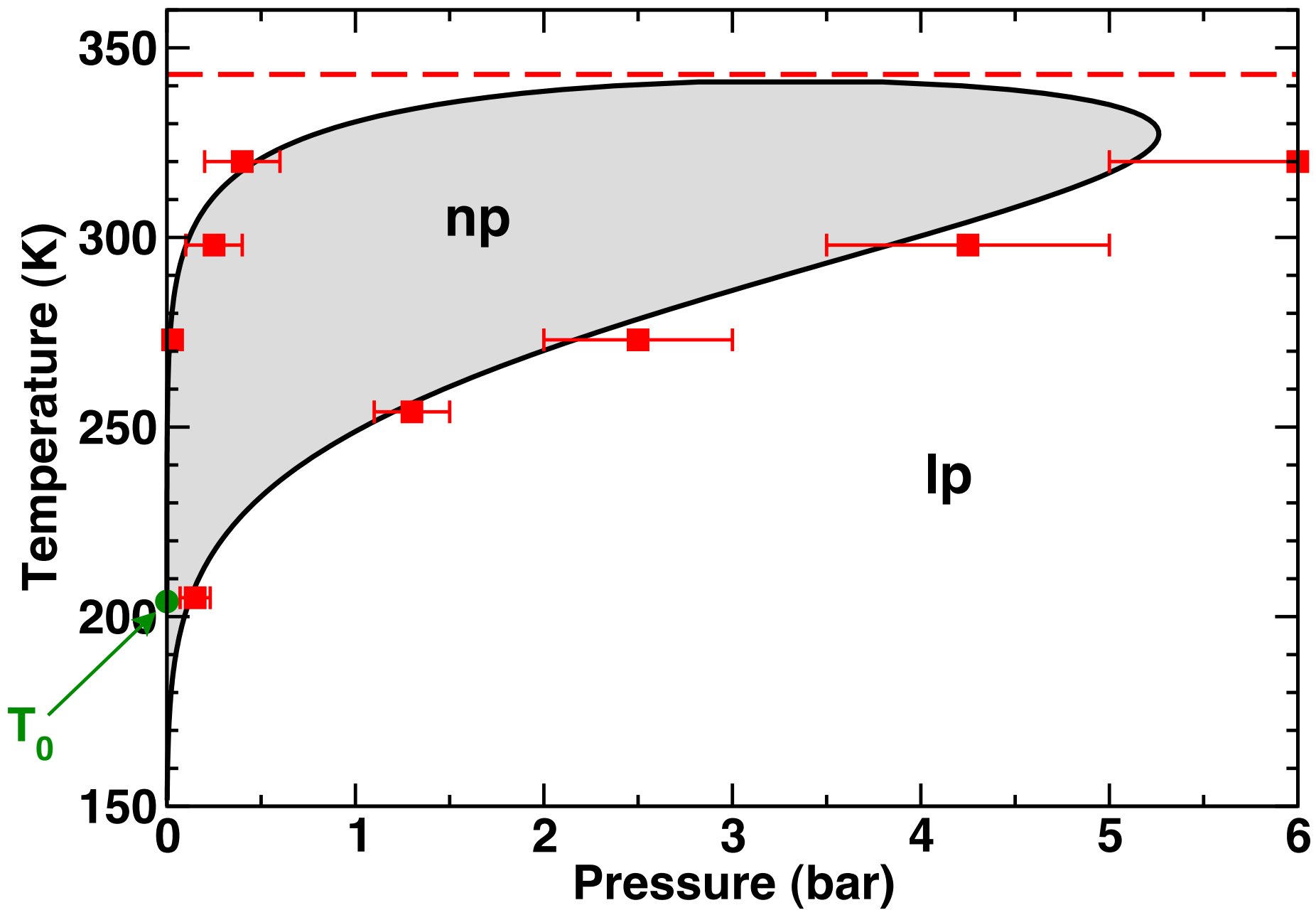

**Figure 6.** Predicted temperature–vapor pressure phase diagram for $CO_2$ adsorption in MIL-53 (Al) material (black line), compared with experimental data points. Red squares (and error bars) represent the observed structural transitions. The dashed red line represents the isotherm at 343 K for which no transition was experimentally observed in this pressure range. $T_0$ is the equilibrium **lp-np** phase transition temperature, derived by our model, in the empty material.

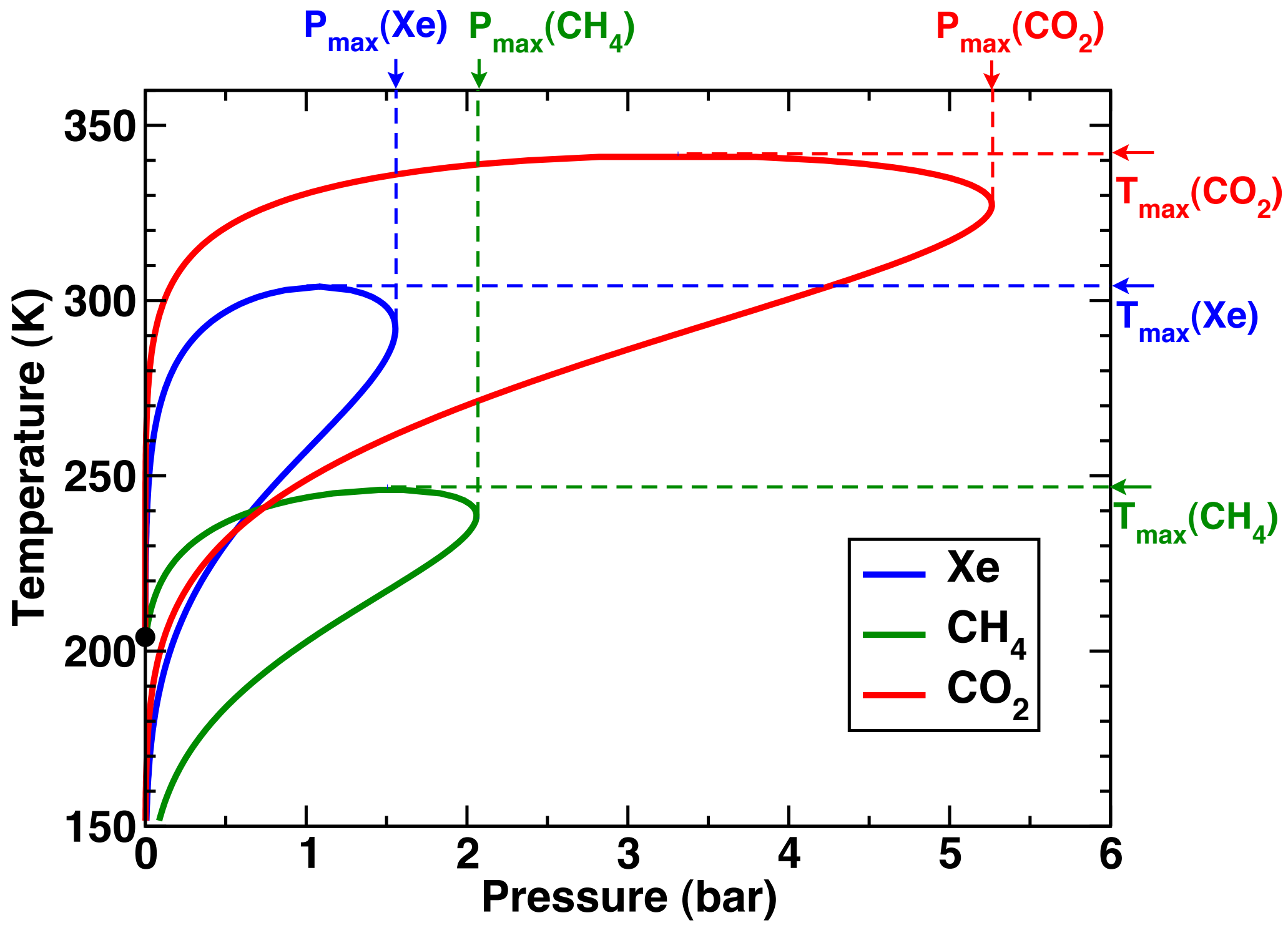

**Figure 7.** Comparison of the temperature–vapor pressure phase diagrams for $CO_2$ (red), $CH_4$ (green) and Xe (blue) adsorption in MIL-53 (Al). The black dot (at P = 0) represents $T_0$, the equilibrium **lp-np** phase transition temperature, derived by our model, in the empty material.

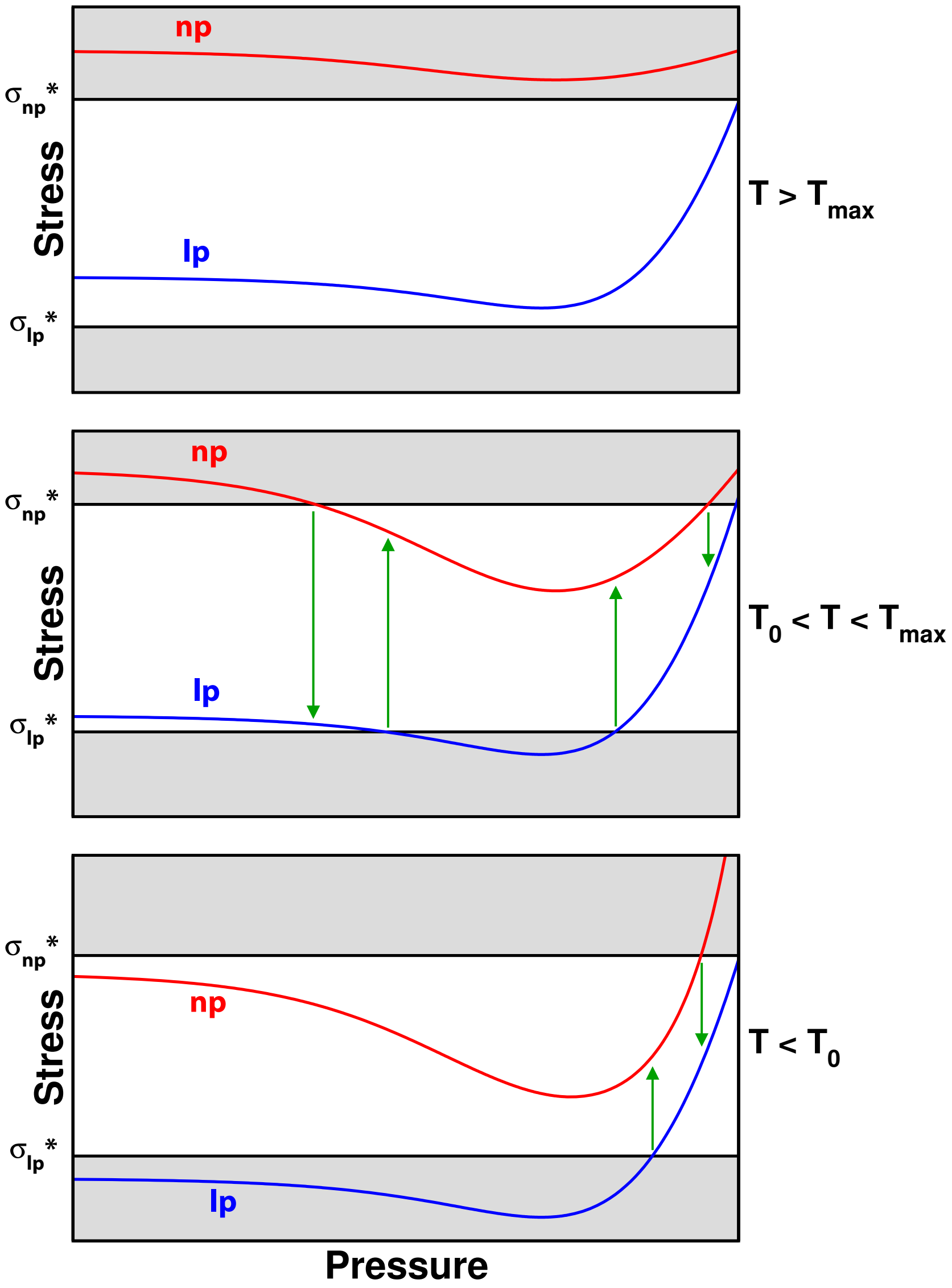

**Figure 8.** Schematic representation of the adsorption stress for both phases (**np**: red; **lp**: blue), and critical stress $\sigma^*_{np}$ and $\sigma^*_{lp}$ determining the structural transitions upon adsorption and desorption (green arrows). $T_{max}$ is the temperature above which no breathing transition takes place for a given guest molecule. $T_0$ is the equilibrium **lp** – **np** transition temperature for the empty material. The 224 K isotherm discussed in the text corresponds to the intermediate panel ($T>T_0$) where the two breathing transitions occur. The lower panel corresponds to the situation in which the stable phase for the empty material is the **np** structure, and thus only one transition is observed.

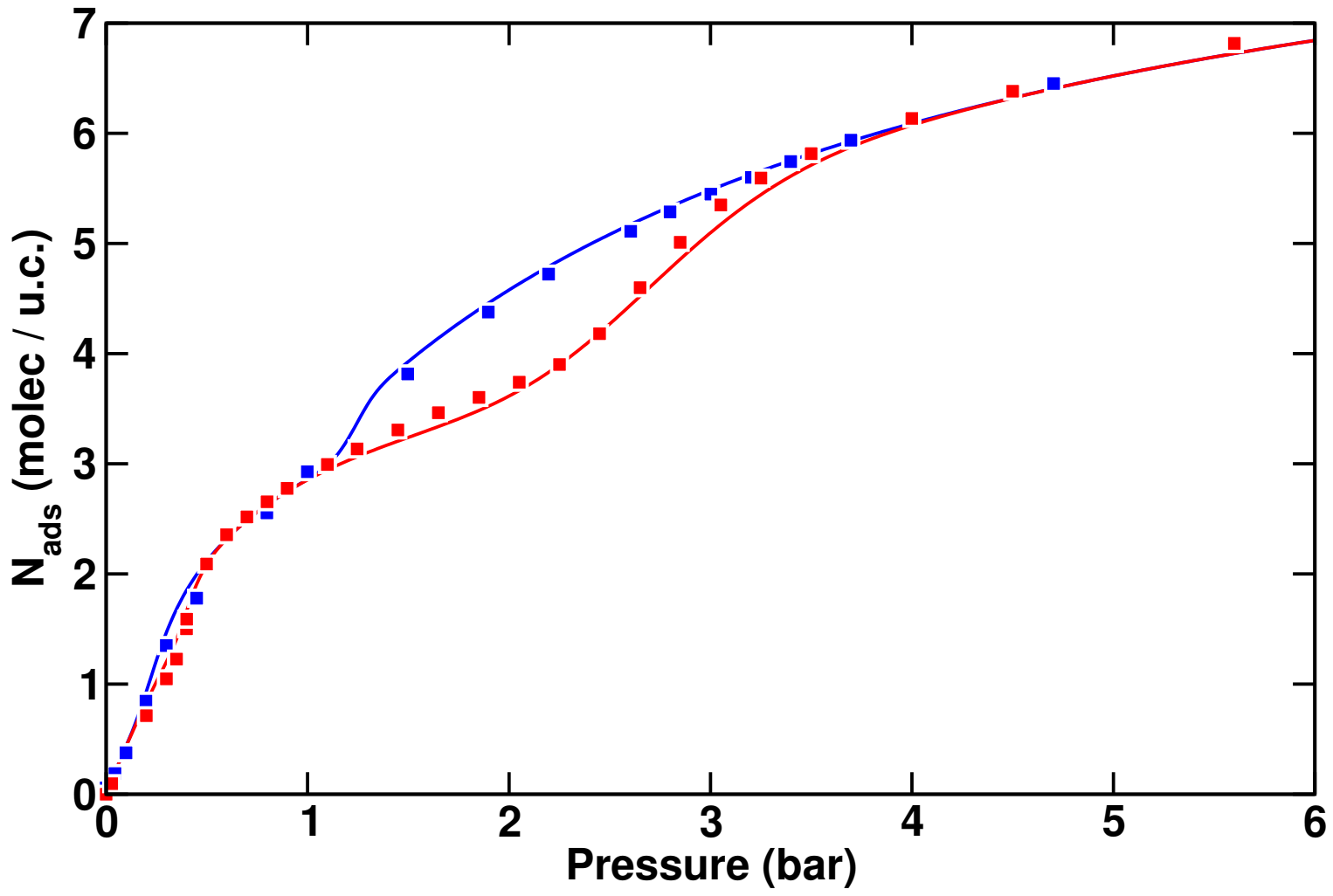

**Figure 9.** Experimental adsorption (red) and desorption (blue) isotherms of $CH_4$ in MIL-53 (Al) at 224 K (square symbols), compared to calculated adsorption and desorption isotherms derived from the stress model (solid lines).

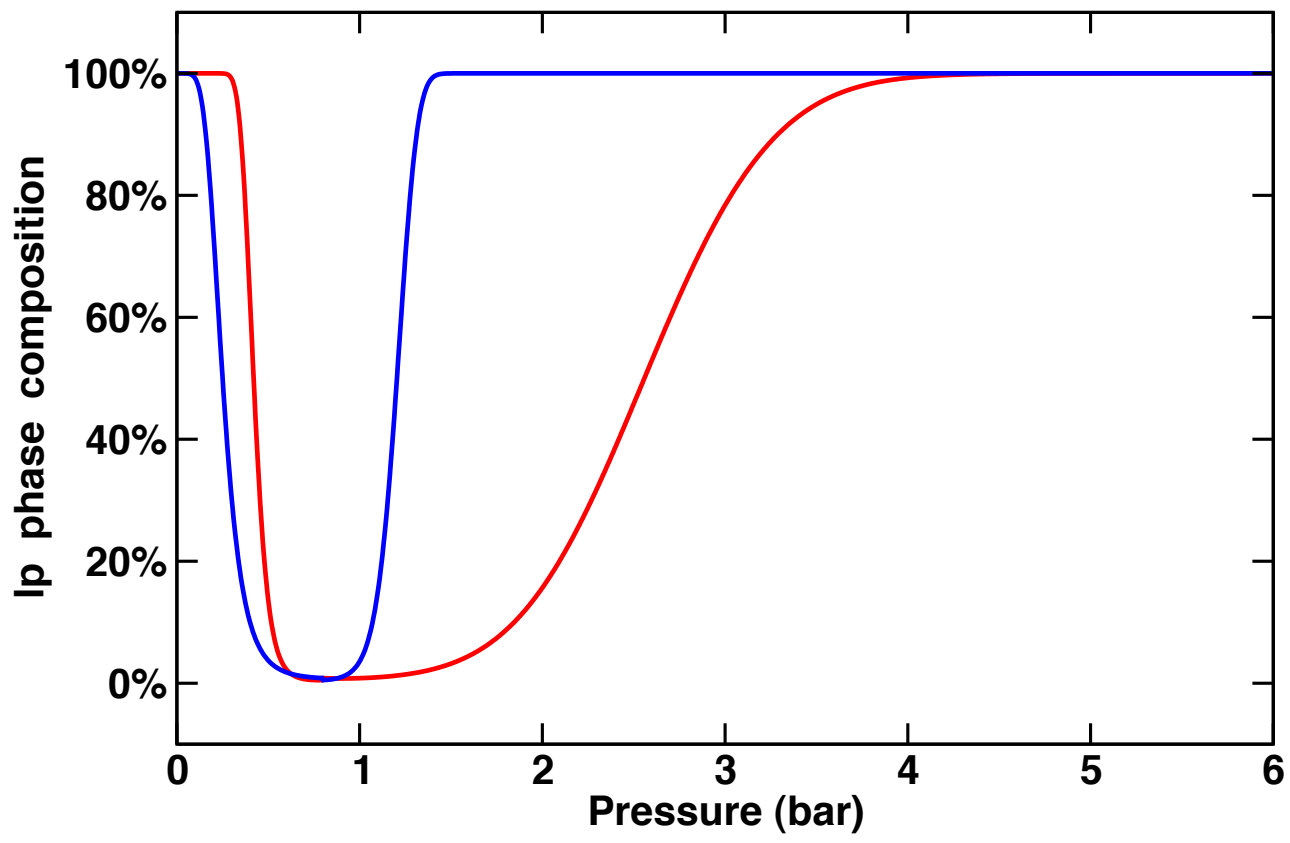

**Figure 10.** Calculated **lp** phase composition for a model representing an assembly of crystallites of different sizes in the stress-based model, upon adsorption (red line) and desorption (blue line) of $CH_4$ in MIL-53 (Al) at 224 K.

| guest | Xe | $CO_2$ | $CH_4$ |
|---|---|---|---|
| $\Delta H_{host}$<br>$\Delta S_{host}$ | 15 kJ/mol<br>74 J/mol/K | | |
| $K_{np}$<br>$\Delta H_{ads}$<br><br>$K'$ | <br>22.2 kJ/mol<br><br>3.0 $10^{-3}$ | <br>38.8 kJ/mol<br><br>1.1 $10^{-5}$ | <br>16 kJ/mol<br><br>1.56 $10^{-3}$ |
| $K_{lp}$<br>$\Delta H_{ads}$<br><br>$K'$ | <br>19.8 kJ/mol<br><br>1.30 $10^{-3}$ | <br>26.0 kJ/mol<br><br>8.5 $10^{-5}$ | <br>15.3 kJ/mol<br><br>1.23 $10^{-3}$ |
| $N_{max,np}$<br>$= a - bT$ | $a = 3.3$<br>$b = 0.002$ | $a = 2.3$<br>$b = 0.0002$ | $a = 4.87$<br>$b = 0.0035$ |
| $N_{max,lp}$<br>$= a - bT$ | $a = 14$<br>$b = 0.02$ | $a = 16.8$<br>$b = 0.025$ | $a = 19.2$<br>$b = 0.045$ |

**Table 1.** Thermodynamic parameters for the temperature–vapor pressure phase diagrams of Xe[18], $CO_2$ and $CH_4$ adsorption in MIL-53 (Al): host free enthalpy ($\Delta H_{host}$) and entropy ($\Delta S_{host}$), temperature dependence of the Langmuir parameters: $N_{max} = a - bT$ in molec/uc; $K = K'$ exp $(-\Delta H_{ads} /RT)$ in molec/uc/bar.

| Host phase | **lp** structure | **np** structure |
|---|---|---|
| $K$ | 4.61 $bar^{-1}$ | 8.40 $bar^{-1}$ |
| $N_{max}$ | 9.09 | 4.32 |
| $(dN_{max}/dV_c)$ | 1.09 $10^{-2}$ Å$^{-3}$ | 1.09 $10^{-2}$ Å$^{-3}$ |
| $(dK/dV_c)$ | –2.17 $10^{-3}$ $bar^{-1}$ Å$^{-3}$ | –3.71 $10^{-2}$ $bar^{-1}$ Å$^{-3}$ |

**Table 2.** Parameters of the stress model for $CH_4$ adsorption in MIL-53 (Al) at 224 K.

**Table of Contents Graphic**

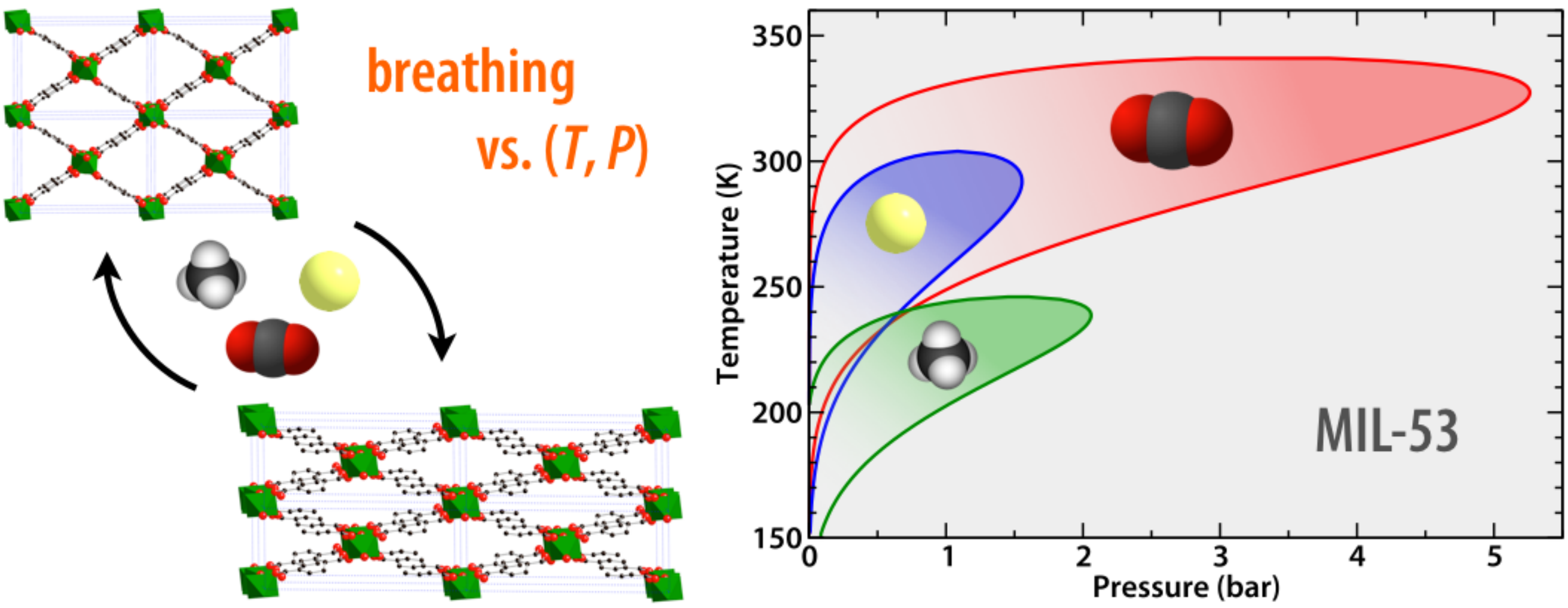